\documentclass[
 aip,
 rsi,
 amsmath,amssymb,
 reprint,
 floatfix
]{revtex4-2}

\usepackage{graphicx}
\usepackage{dcolumn}
\usepackage{bm}

\usepackage[utf8]{inputenc}
\usepackage[T1]{fontenc}
\usepackage{mathptmx}
\usepackage{etoolbox}
\usepackage{placeins}

\makeatletter
\def\@email#1#2{%
 \endgroup
 \patchcmd{\titleblock@produce}
  {\frontmatter@RRAPformat}
  {\frontmatter@RRAPformat{\produce@RRAP{*#1\href{mailto:#2}{#2}}}\frontmatter@RRAPformat}
  {}{}
}%
\makeatother
\begin{document}

\preprint{AIP/123-QED}

\title[]{A comprehensive semi-automated fabrication system for quartz tuning fork AFM probe with real-time resonance frequency monitoring and Q-factor control}

\author{Hankyul Koh}
 \email{physics113@snu.ac.kr.}
\affiliation{ 
Physics, College of Natural Sciences Dept. of Physics \& Astronomy, Seoul National University, South Korea
}%
\author{Joon-Hyuk Ko}
\affiliation{ 
Center for AI and Natural Sciences, Korea Institute for Advanced Study, South Korea
}%
\author{Wonho Jhe}
\affiliation{%
Multiscale Instruments, Seoul, South Korea
}%




\date{\today}
             
\begin{abstract}
Quartz tuning fork–based atomic force microscopy (QTF-AFM) has become a powerful tool for high-resolution imaging of both conductive and insulating samples, including semiconductor structures and metal-coated surfaces as well as soft matter under ambient conditions, while also enabling measurements in more demanding environments including ultrahigh vacuum and cryogenic conditions where conventional cantilever-based AFM often encounters limitations. However, the broader adoption of QTF-AFM has been constrained by the difficulty of attaching a cantilever tip to a quartz tuning fork (QTF) with the positional and angular precision required for repeatable and reproducible probe fabrication. For stable operation, the tip must be placed precisely at the midline of a single tine, aligned parallel to the prong axis, and rigidly secured. Even slight lateral offsets or angular deviations disrupt the intrinsic antisymmetric flexural mode, induce torsional coupling, and ultimately lead to systematic image distortions and reduced measurement integrity. In this work, we present a comprehensive, semi-automated QTF-tip fabrication system that integrates precision alignment, real-time frequency-sweep monitoring, and controlled Q-factor tuning within a single workflow. Experimental characterization demonstrates consistent probe preparation across multiple trials, preservation of sharp and well-defined resonance responses with deliberately adjustable damping, and high-fidelity, high-resolution imaging in practical scanning tests. This integrated approach provides a reproducible framework to QTF-based probe fabrication, lowering the technical barrier to QTF-AFM implementation and broadening its applicability across diverse sample types and operating environments.
\end{abstract} 

\maketitle

\section{\label{sec:level1}Introduction}
Atomic force microscopy (AFM) \cite{Binnig1986_AFM} has become an essential tool for nanoscale surface characterization across physics, materials science, and the semiconductor industry. While conventional microcantilever-based AFM enables high-resolution imaging in ambient conditions, its performance degrades in environments requiring mechanical stability, low thermal drift, or operation in ultrahigh vacuum (UHV) and cryogenic temperatures \cite{Meyer1988_OpticalLeverAFM,Marti1992_LaserIrradiationCantilevers, Thundat1994_ThermalAmbientDeflections, Radmacher1995_ImprovementThermalBending, Wenzler1996_ImprovementsAFMCantileversStability,Hug1999_LowTempUHV_SFM,Suehira2000_LT_UHV_OBD_AFM,Giessibl2011_ComparisonForceSensors_QTF_LER}. In such regimes, quartz tuning forks (QTFs) offer distinct advantages over cantilever probes due to their high stiffness, self-sensing capability, minimal thermal drift, and compatibility with multifunctional measurement  \cite{FriedtIntroductionQuartzTuning2007,AbrahamiansContributedReviewQuartz2016,Giessibl2019_qPlusSensor}. As a result, QTF-AFM \cite{GuntherFischerDransfeld1989SNAM,Karrai1995_PiezoTipControl,Giessibl2003RMP,Naber1999_dynamic} has enabled high-resolution imaging in UHV, low-temperature, and other demanding environments, as well as reliable scanning of soft or fragile materials under ambient conditions \cite{Karrai1995_PiezoTipControl,Giessibl2003RMP}. 

Despite these advantages, the widespread use of QTF-AFM remains limited by a central and long-standing bottleneck: the proper fabrication of a QTF–tip assembly \cite{Naber1999_dynamic}. Attaching a microscale tip to a QTF prong with the positional and angular tolerances required for stable operation is technically challenging, and small errors readily induce torsional coupling \cite{Tung2010_AngledTip}, degrade the resonance characteristics, and compromise imaging fidelity. This difficulty has discouraged many researchers from adopting QTF-based methods. Leveraging microfabricated commercial AFM probes offers a practical route to improved and standardized tip sharpness (nominal radii down to about $2~nm - 10~nm$), as well as well-characterized geometries and readily available functionalizations, compared with in-house tips fabricated by electrochemical etching, focused-ion-beam (FIB) micro-welding, or pulled-fiber rod methods, whose apex shape and radius can vary substantially from probe to probe \cite{ Grundner1986_HydrophobicSilicon,JerschMaletzkyFuchs2006InterfaceCircuits,Karrai1995_PiezoTipControl,Giessibl1998Sensor,SalviEtAl1998ShearForceDetection,Naber1999_TF_SNOM}. 

Another challenge with fabricating a proper QTF-AFM probe is related to the intrinsically high quality factor of the QTF sensor, which is typically $10^{4}-10^{5}$ in vacuum and $10^{3}-10^{4}$ under ambient conditions \cite{Grober2000_QTF_limits}. Such high Q-values are incompatible with amplitude-modulation AFM (AM-AFM) in air, where the effective response time $\tau = 2Q/\omega_{0}$ limits the usable bandwidth to only a few hertz \cite{Albrecht1991FM, Garcia2002DynamicAFM}. Practical AFM imaging therefore requires reducing the Q-factor to the range of several hundreds. A common strategy is to introduce mechanical asymmetry by adding mass to one prong of the fork, thereby lowering the Q-factor \cite{Cleveland1993AFM,Kim2014EffectiveStiffness}. An extreme implementation of this concept is the qPlus configuration \cite{Giessibl2019_qPlusSensor}, in which one tine is rigidly immobilized and the remaining tine behaves as a single-prong cantilever. Although qPlus sensors are widely used, they provide limited flexibility in tip replacement and offer virtually no means for deliberate Q-factor tuning. Alternative approaches such as mechanically shortening one tine or attaching beaded weights \cite{HussainOptimizingQualityFactor2017} can, in principle, modify the Q-factor. However, these methods suffer from poor reproducibility, large sample-to-sample mass variation, uncontrolled attachment geometry, and the introduction of suboptimal resonance characteristics.

Here, we employ a Q-factor adjustment method based on a liquid adhesive, which provides significant advantages from this perspective. Because a liquid naturally minimizes its surface area for a given volume following the principle of least action, it spontaneously achieves geometrical symmetry. Thus, once the cantilever tip is attached precisely at the central thickness of one prong, the adhesive layer preserves symmetry without distortion. Therefore, the adhesive-based method presented here is consistently yields sharp, symmetric resonance curves with minimal mode distortion. Moreover, the approach introduced here is compatible with a wide variety of commercially available cantilever tips --- including silicon, silicon nitride, and electron-beam-deposited (EBD) tips --- allowing researchers to freely select or replace the tip according to experimental needs while retaining full control over the dynamical properties of QTFs.

In this work, we introduce a rapid, reliable, and robust method for attaching a cantilever tip to a standard two-prong quartz tuning fork. The procedure offers high positional accuracy, ensures excellent repeatability, and allows controlled tuning of the Q-factor through precise adjustment of adhesive volume. By continuously monitoring the resonance characteristics of the QTF throughout assembly, the method preserves the intended flexural dynamics and prevents the accumulation of mechanical asymmetries. This approach substantially lowers the technical barrier to QTF-AFM and provides a practical and reproducible route for preparing customizable QTF probes suitable for a wide range of experimental environments.

\begin{figure}[tbp]
\includegraphics[width=0.45\textwidth]{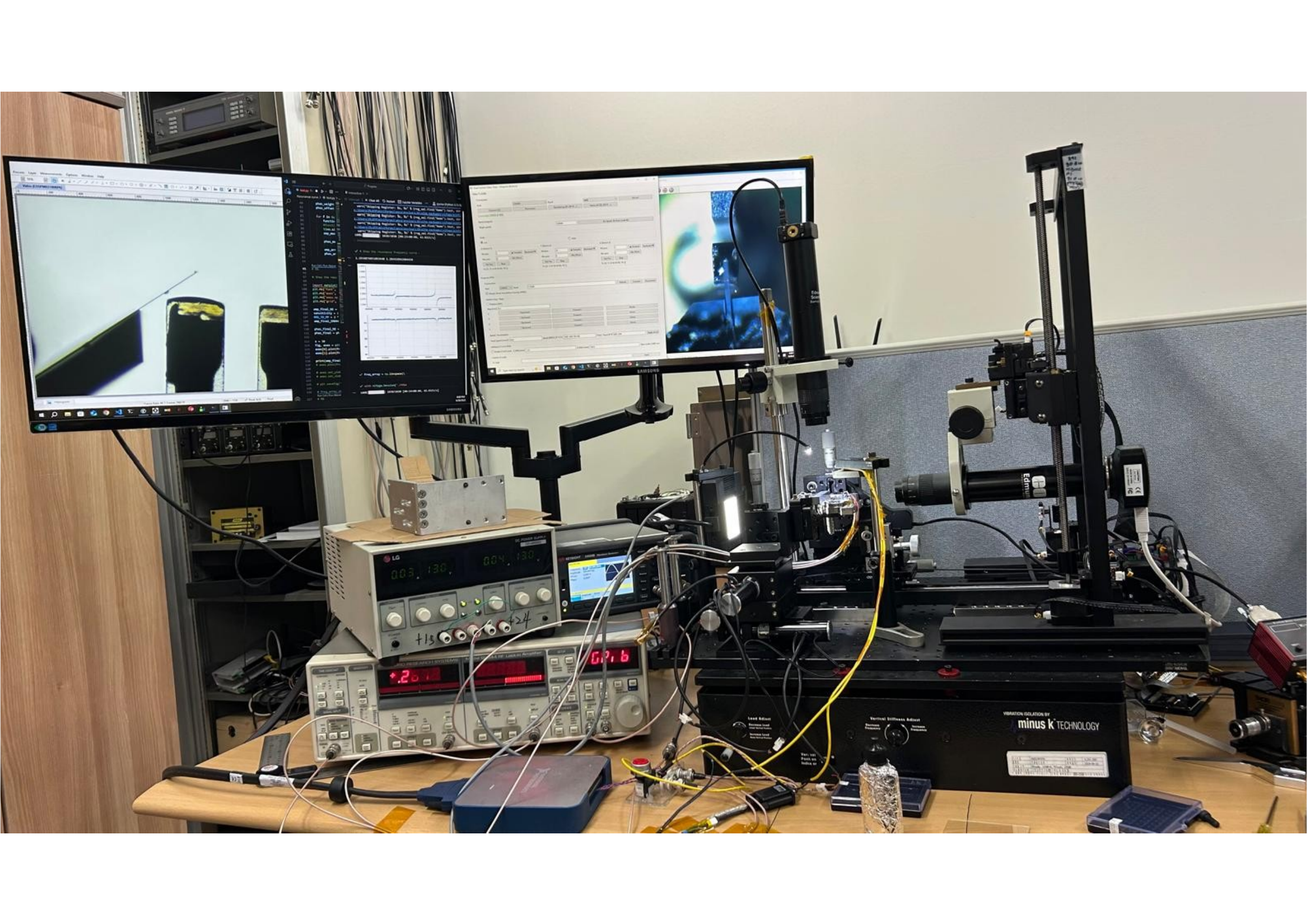}
\caption{\label{fig:setup}  Overall view of the experimental setup used for QTF-AFM probe fabrication and characterization, including the arbitrary waveform generator, lock-in amplifier, transimpedance amplifier, FPGA-based acquisition system, and supporting illumination optics and three-axis linear actuators for precise positioning.}
\end{figure}


\section{System Overview}
Accurate imaging with QTF-AFM requires stabilizing several coupled mechanical and electronic parameters. 
The most critical requirement is that the resonance response near the operating mode be reproducibly described by a well-defined, symmetric Lorentzian profile, indicating minimal nonlinear distortion and a high signal-to-noise ratio in the frequency-amplitude spectrum. 
Mechanically, QTF-tip fabrication must avoid morphological asymmetry: when a mass is attached to one prong, the added tip should remain centered about the tine thickness axis (side view) to prevent lateral imbalance and undesired torsional coupling. 
Electronically, reliable resonance characterization requires suppressing stray-capacitance feedthrough and maintaining a flat transimpedance gain over the relevant frequency range. 
In this work, the deposited adhesive volume serves as a practical control parameter to tune the Q-factor from the bare value in air ($Q\approx 2000$) to application-specific ranges (typically $Q\approx 100-700$ for high-speed imaging and $Q\approx 700-1200$ for high-resolution imaging), while preserving Lorentzian reproducibility through a compensated, low-noise transimpedance front end.

\begin{figure}[tbp]
\includegraphics[width=0.43\textwidth]{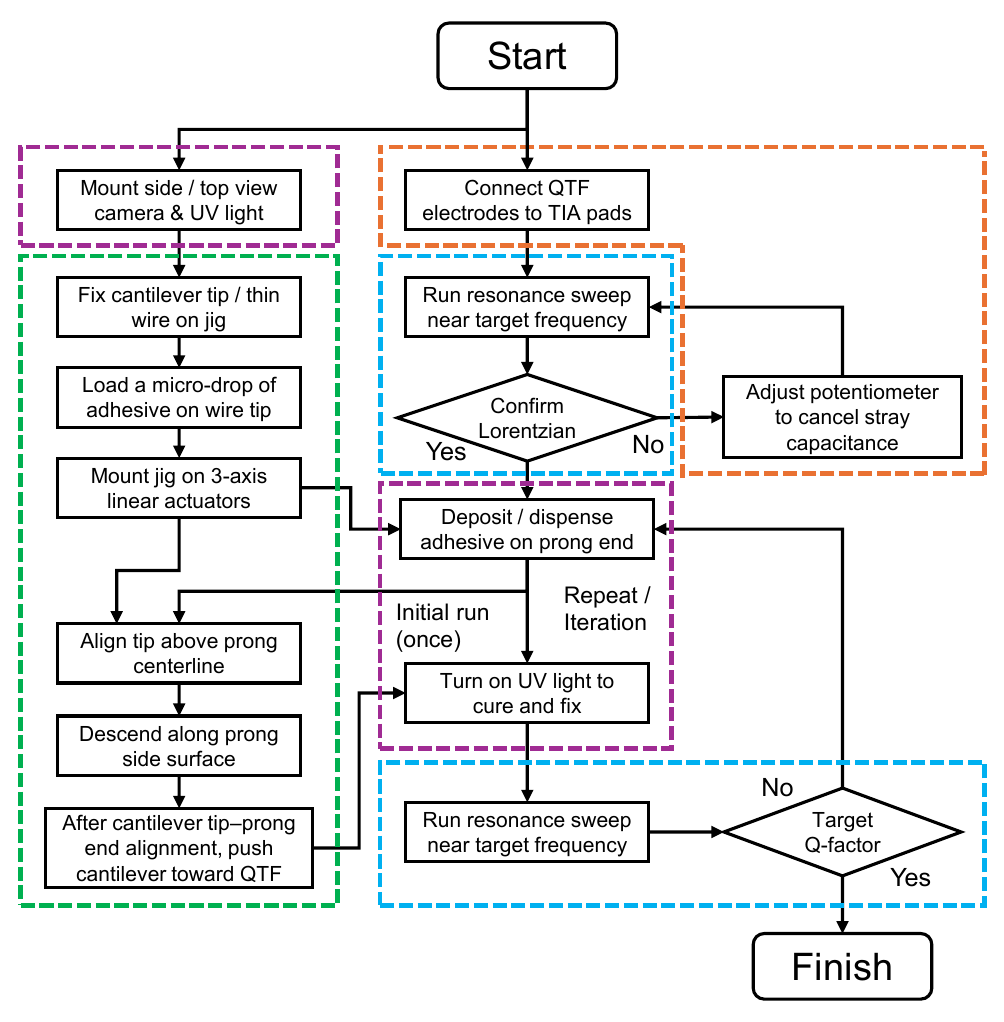}
\caption{\label{fig:flow} Flow chart of the QTF-tip fabrication, frequency monitoring, and Q-factor control system.}
\end{figure}



\section{Experimental Setup}
The experimental system is designed for precise cantilever-tip attachment onto a QTF with real-time resonance characterization and controlled tuning of the Q-factor. The setup is organized into four subsystems: (i) the QTF-tip assembly, (ii) electronics, (iii) motion-control and alignment subsystem, and (iv) data acquisition and processing subsystem. We provide details for each component below.

\subsection{QTF-tip assembly}
The QTF was mounted on a holder equipped with complementary CMOS cameras providing both top and side views for precise visual alignment. For the top-view imaging, the camera was intentionally tilted by approximately $30^\circ-45^\circ$ relative to the horizontal plane to simultaneously visualize lateral alignment between the cantilever tip and the side face of the QTF prong.
The QTF mount was placed on a pitch-yaw adjustment plate, enabling accurate control of both horizontal and vertical alignment between the QTF prong and the cantilever tip.

\subsection{Electronics}
The electronics subsystem was comprised of an arbitrary waveform generator (33500B, Keysight), a lock-in amplifier (SR844, Stanford Research Systems), a data-acquisition board (NI USB-7855R, National Instruments) and a custom-built transimpedance amplifier (TIA) optimized for low input-bias operation and wide bandwidth.

Of the two electrodes of the QTF, the drive electrode was connected to the arbitrary waveform generator, which supplied the voltage drive signal to actuate the QTF. The remaining sensing electrode was connected to the TIA, which amplified and converted the self-sensing current response of the QTF into a voltage signal, which was then routed to the lock-in amplifier \emph{Signal In}. For phase-referenced detection, the synchronization output (\emph{Sync}) of the waveform generator was connected to the lock-in \emph{Ref In}. Subsequently, the lock-in in-phase and quadrature outputs ($X$ and $Y$) were sent to the analog in channels of the data acquisiton board for real-time logging.

All electrical connections were made using shielded $50~\Omega$ BNC coaxial cables to minimize pickup and ensure reproducible wiring capacitance.
As the signal current generated by a QTF is typically at the pA level \cite{Karrai1995_PiezoTipControl} and resonance monitoring and feedback can be performed at higher flexural modes, typically $5-6$ times the fundamental frequency \cite{Kim2017_EigenmodesQTF}, the TIA must provide a low-noise front end with sufficiently wide closed-loop bandwidth so that the transimpedance gain does not roll off appreciably.
Such stable operation was ensured by minimizing the physical separation between the QTF and the TIA input, adding a properly chosen feedback capacitor in parallel with the feedback resistor for phase compensation, and incorporating a dummy load that emulates the sensor shunt capacitance.
This balancing approach suppressed resonance-like artifacts and feed-through associated with stray capacitances, improving robustness and reproducibility of the Lorentzian resonance spectra \cite{An2013_SCC}.

\subsection{Motion control and alignment subsystem}
The motion-control subsystem consisted of three daisy-chained linear actuators (T-LA28A, Zaber) and associated stages for cantilever positioning and adhesive handling. All three translational axes ($x$, $y$, and $z$) were operated in closed loop using the built-in encoders.
A Python-based motion-control program was developed to communicate with the motor controllers via RS232 and to execute deterministic positioning commands with high precision with a minimum step size of $0.1~\mu m$ for alignment and attachment procedures.

A custom-designed jig was mounted on an intermediate rotational stage allowing rotation about the longitudinal axis of the cantilever, enabling precise angular alignment of the cantilever with respect to the QTF prong prior to attachment.
The jig incorporated two independent holders.
One holder carries a fine wire for contact-mediated micro-deposition, also called dabbing, of the adhesive onto the QTF surface, and the other holder supports the cantilever tip for subsequent placement.
Because controlling an nL-scale adhesive volume is nontrivial, even with microfluidic dispensers where viscosity and surface tension can hinder accurate dosing at the outlet, the dabbing transfer provides a reliable and repeatable method for small-volume deposition.
As an alternative, a piezo-actuated micro-injection device capable of dispensing approximately $3-5~nL$ may also be employed.
Together, the system provides two rotational degrees of freedom at the QTF mount, pitch and yaw, and four degrees of freedom at the cantilever stage, three translations and one axial rotation, yielding full six-degree-of-freedom control of the relative alignment.

Real-time visual feedback of the top and side views with an optional front view was provided using CMOS cameras, a 365~nm, 3~W ultraviolet LED source was used for adhesive curing. 
All motion stages and measurement instruments are controlled using Python-based software, enabling automated resonance-frequency sweeps and synchronized actuation during tip attachment.

\subsection{Data acquisition and processing subsystem}
The QTF is driven by sinusoidal voltage generated by an arbitrary waveform generator (33500B, Keysight), and the piezoelectric response current is converted to a voltage by the custom TIA and demodulated by a lock-in amplifier (SR844, Stanford Research Systems).
The drive amplitude is typically set to $0.2-2.0~\mathrm{V_{pp}}$.
Instrument control and measurement automation are implemented via the GPIB interface using \texttt{PyVISA}.
The in-phase and quadrature outputs of the lock-in ($X$ and $Y$) are acquired as analog voltages using the analog input channels of an NI USB-7855R (National Instruments).
The device is operated with a precompiled LabVIEW FPGA bitfile, and the sampled $X$ and $Y$ signals are streamed to the host computer through the \texttt{nifpga} Python interface.
The lock-in sensitivity, time constant, and roll-off are selected to ensure a sufficiently narrow equivalent noise bandwidth.
At each frequency, the steady-state $X$ and $Y$ voltages are recorded and converted to response amplitude and phase, calibrated using independently determined scaling coefficients, and expressed as an equivalent input current.
The resulting amplitude and unwrapped phase spectra are fitted using the analytical model of Lee \emph{et al.} \cite{Lee2007_QTFquantitative}, which describes an electrically driven tuning fork as a parallel $LRC$ branch in parallel with a stray capacitance $C_0$.
From these fits, the resonance frequency and Q-factor are extracted directly from the frequency-domain current response.

\begin{figure}[tbp]
\includegraphics[width=0.45\textwidth]{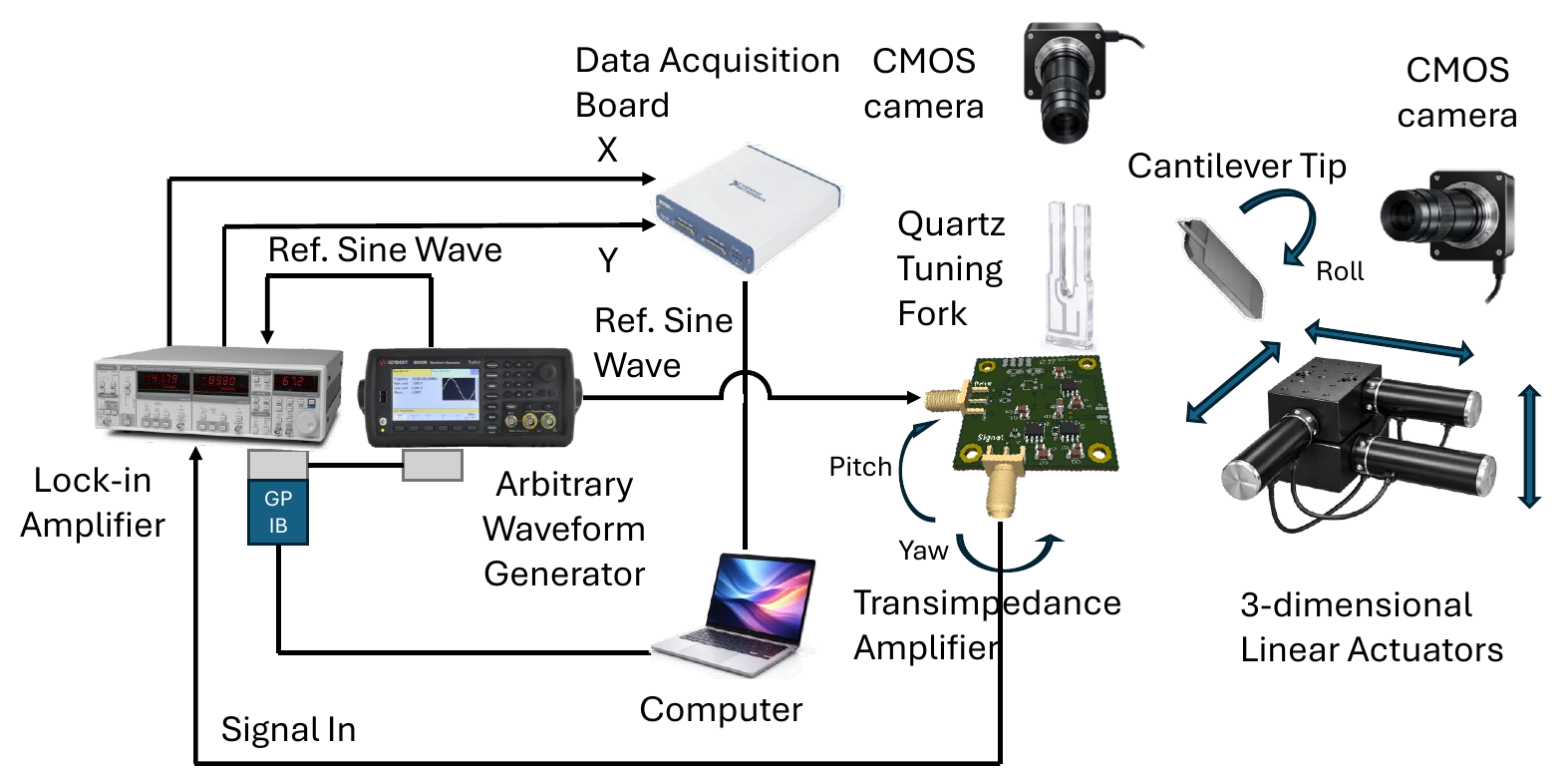}
\caption{\label{fig:overall diagram} Schematic diagram of the overall fabrication system.}
\end{figure}


\vspace{-5pt}

\section{Methods}

\begin{figure}[!ht]
\includegraphics[width=0.45\textwidth]{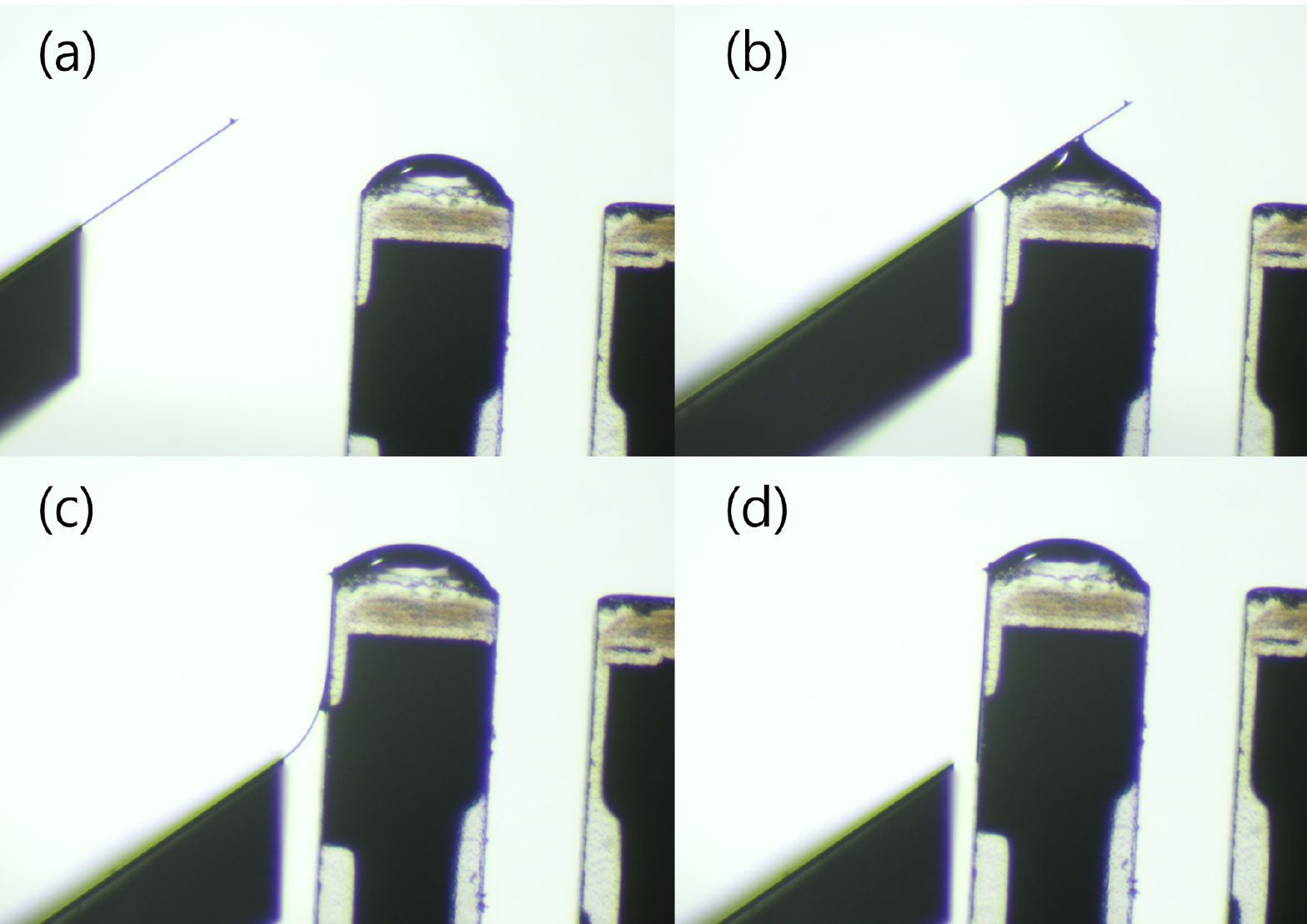}
{\caption{\label{fig:tip_fabrication}
Step-by-step fabrication process of a QTF-based AFM tip. (a) A sharp AFM cantilever tip is precisely aligned with the end facet of a QTF prong. (b) The tip is brought into controlled mechanical contact with the prong apex at a defined approach angle. (c) A minute amount of adhesive forms a junction between the tip and the QTF prong, stabilized by capillary forces arising from the adhesive meniscus. (d) The tip is aligned parallel to the QTF prong and advanced in $5~\mu m$ steps using a linear actuator; immediately upon fracture, the adhesive is UV-cured to fix the tip in place.}}
\end{figure}

After mounting the QTF and preparing the cantilever tip on the positioning jig, an initial frequency sweep is performed to confirm a symmetric Lorentzian response and to determine the baseline resonance frequency and quality factor.
A small droplet of ultraviolet-curable adhesive is then placed on the end of the selected prong, and the deposited adhesive volume is typically in the range of approximately $2.5-25~nL$.
To achieve reliable attachment, the adhesive and the attachment process satisfy two requirements.
First, the adhesive has sufficiently low density and viscosity so that surface forces dominate competing contributions, including the elastic restoring force of the deflected cantilever, enabling robust attraction to the mechanically compliant silicon cantilever.
Second, the adhesive supports rapid ultraviolet curing with sufficient rigidity to preserve alignment after fixation.
The exposed cross-section at the end of the QTF prong is intrinsically hydrophilic, which promotes adsorption and retention of the adhesive.
Although silicon is intrinsically hydrophobic, under ambient conditions it is covered by an adsorbed water layer \cite{Grundner1986_HydrophobicSilicon, Xi2010_AFM_SiliconSurface, Israelachvili2011_ISF, Chen_WaterAdsorption_Silicon}.
Together with van der Waals forces, this can produce strong attractive interactions at small separations that draw the cantilever underside into contact with the adhesive and maintain robust anchoring.
Because silicon remains barely wetting, the adhesive remains confined to the droplet surface even when the cantilever underside is firmly attached, so lateral translation by the linear actuators does not induce capillary spreading along the cantilever sidewalls and helps avoid unintended wetting and contamination of the tip.
During alignment, the actuators translate the cantilever laterally in discrete steps of $0.5-5.0~\mu m$, allowing the underside of the cantilever to slide controllably along the droplet surface toward the contact-line region.
This guided translation enables alignment along the prong sidewall while maintaining a symmetric adhesive meniscus about the tine thickness, suppressing lateral imbalance and minimizing torsional coupling.
Once a stable meniscus forms between the cantilever underside and the prong sidewall, ultraviolet illumination at 365~nm is applied to cure the adhesive and preserve the final alignment.
The attached cantilever position and orientation are verified by optical inspection, after which a frequency sweep is performed to quantify the resonance shift and the resulting Q-factor following attachment.

If the measured Q-factor does not match the target value, additional adhesive is deposited in controlled increments, followed by UV curing, and the resonance spectrum is re-measured after each step.
This deposit-cure-sweep cycle is repeated until the desired operating range is reached.
Near the target, a smaller final increment is used to minimize overshoot and achieve reproducible Q-factor tuning.

\begin{figure}[!ht]
\includegraphics[width=0.45\textwidth]{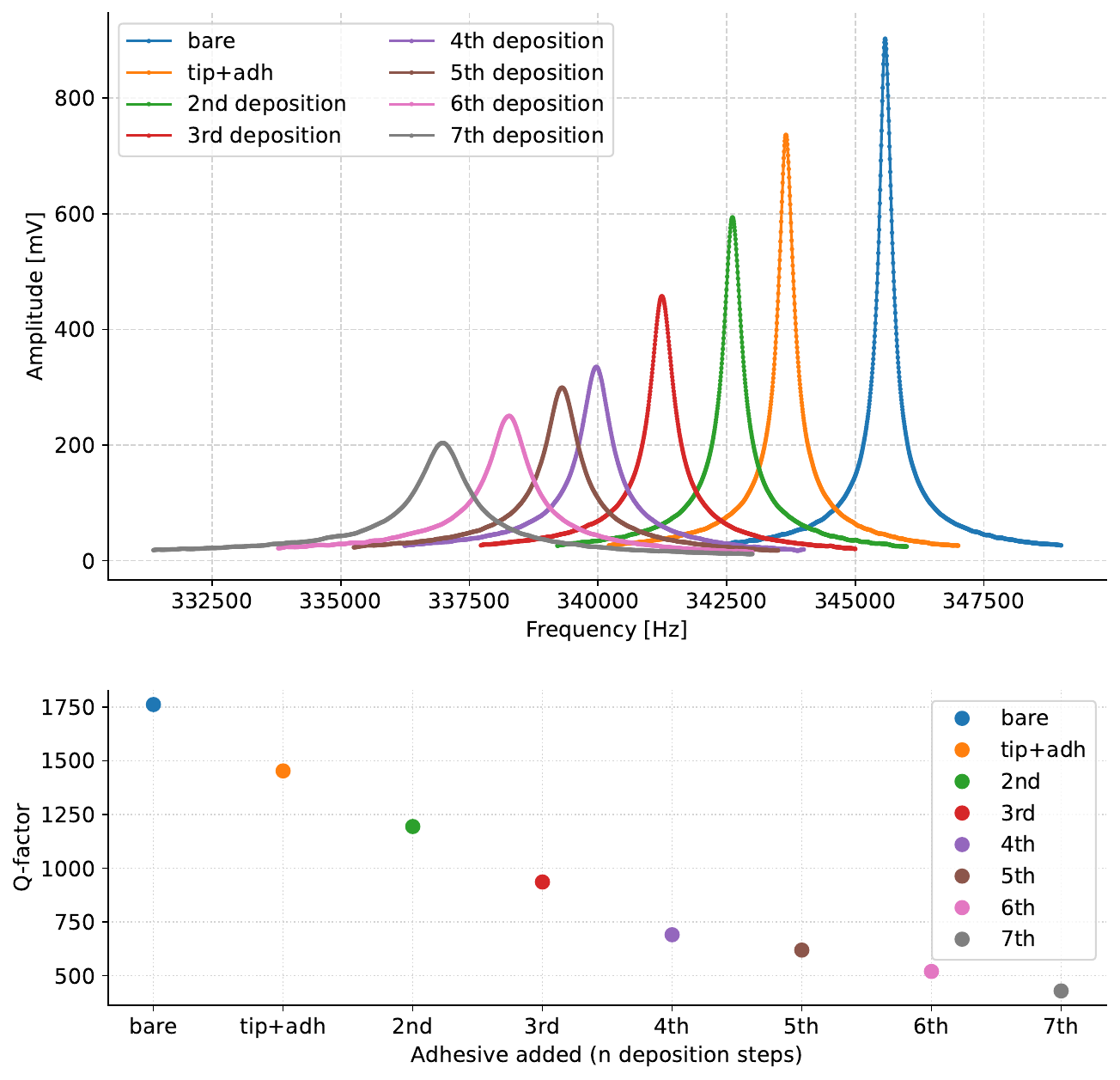}
\caption{\label{fig:noa_q}
Resonance spectra of the second flexural mode for each adhesive-deposition step, yielding the resonance frequency $f_0$ and amplitude at each step \textbf{(Top)}, where the added adhesive mass per step is approximately $2.5~\mu g$.
Corresponding Q-factors extracted for each deposition step \textbf{(Bottom)}.
}
\end{figure}

\section{Experiments}
The fabrication workflow proposed in this paper is experimentally validated through three complementary measurements. First, Fig.~\ref{fig:tip_fabrication} demonstrates that the attachment can be executed in a deterministic manner while preserving the geometric symmetry required to maintain the intended antisymmetric flexural dynamics, thereby suppressing lateral imbalance that would otherwise excite torsional coupling. Second, the resonance measurements in Fig.~\ref{fig:noa_q} verify that nanoliter-scale adhesive loading provides a quantitative tuning control parameter for the QTF dynamical response. The monotonic evolution of $f_0$ and the systematically reduced $Q$ extracted from Lorentzian fits confirm controlled modification of dissipation, and the persistence of symmetric curves indicates that the tuning does not introduce significant mode distortion.

\begin{figure}[tbp]
\includegraphics[width=0.45\textwidth]{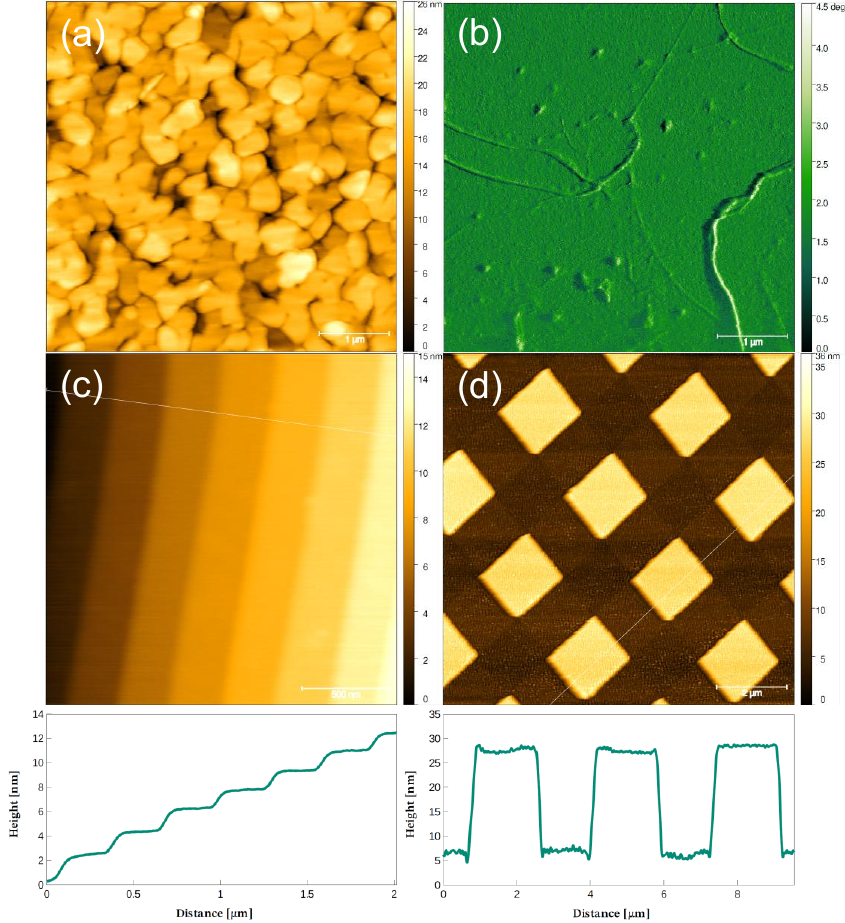}
\caption{\label{fig:qtf_afm_images}
AFM images obtained with a home-made quartz tuning fork (QTF)-based AFM.
(a) Atomic terraces of a Au~(111) thin film grown on a mica substrate.
(b) Carbon nanotubes (CNTs) deposited on a mica substrate.
(c) Commercial SiC calibration sample (TipsNano) featuring a single step height of $1.5~nm$.
(d) Also commercial periodic test structure calibration sample TGQ1 (TipsNano) with a nominal step height of $20 \pm 3~nm$.
All images were acquired using QTF-tip sensors fabricated in this work from SSS-FM (Nanosensors) and PNP-DB (Nanoworld) cantilever tips following the procedure described in this paper.
The measurements were performed using fully home-built QTF-AFM control electronics and software.
}
\end{figure}

Finally, the scanning tests in Fig.~\ref{fig:qtf_afm_images} confirm that the fabricated QTF-tip sensors operate reliably in practical AFM measurements. The clear reproduction of atomically stepped samples and calibrated nanometer-scale step heights without systematic artifacts demonstrates that the combined alignment, resonance monitoring, and Q-factor tuning constitute a robust and reproducible route to QTF-tip fabrication.

\section{Conclusion}
This work addresses a central practical limitation that has constrained the broader adoption of QTF-AFM, specifically the difficulty of reproducibly attaching a sharp cantilever tip to a standard two-prong quartz tuning fork with sufficient positional and angular accuracy to preserve the intended antisymmetric flexural dynamics.
The main conclusion is that sub-micrometer, encoder-based closed-loop positioning enables deterministic QTF-tip fabrication while simultaneously providing quantitative control of the QTF dynamical response through nanoliter-scale adhesive loading.
By adjusting the deposited adhesive volume during assembly, the Q-factor can be tuned as a deliberate fabrication parameter rather than an uncontrolled byproduct of tip attachment.
Real-time frequency-sweep monitoring of the eigenmodes provides immediate feedback on the resonance frequency, peak amplitude, and extracted Q-factor, allowing the deposition and curing steps to be iterated until a target operating regime is reached.
In this way, the workflow enables fabrication of QTF-tip sensors with a user-defined Q-factor, one of the most important parameters governing bandwidth, stability, and imaging performance such as response speed and force sensitivity in QTF-AFM, while maintaining reproducible and symmetric resonance responses within the desired frequency range.
Beyond resonance tuning, the same fabrication and control workflow enables reliable preparation of QTF-tip sensors from commercial silicon cantilevers using an alignment procedure that minimizes lateral imbalance about the tine thickness axis. The resulting sensors, together with fully home-built QTF-AFM were used to obtain representative AFM images across multiple test samples, including Au~(111) on mica, carbon nanotubes on mica, and calibrated step-height standards. These results demonstrate that the proposed combination of mechanically symmetric attachment, controlled Q-factor tuning, and stable electrical readout provides a compact and reproducible route to QTF-AFM instrumentation suitable for both high-speed operation, where reduced $Q$ improves bandwidth, and high-resolution imaging, where higher $Q$ can be maintained to maximize force sensitivity.
Future work will focus on quantifying the relationship between deposited adhesive volume, added mass distribution, and the resulting changes in modal parameters, including resonance frequency, effective stiffness, and mode-dependent dissipation. 


\begin{acknowledgments}
This work was supported by grants from the National Research Foundation of Korea (No. 2016R1A3B1908660) and from the Ministry of SMEs and Startups, Korea, under the 2024 Startup Leading University Support Program (No. 20248082), the 2025 Startup Leading University Support Program (No. 20319065), and under the DeepTech TIPS Program (No. RS-2025-25461650). J. H. Ko was also supported by the KIAS Individual Grant (AP102401) from the Korea Institute for Advanced Study. The authors thank J. H. Seo and J. W. Yun of Multiscale Instruments for their assistance with the experiments.
\end{acknowledgments}

\bibliographystyle{apsrev4-1}
\bibliography{references}

\end{document}